\documentclass[twocolumn]{aastex631}

\usepackage{lipsum} 
\usepackage{tabularx}
\usepackage{amsmath}
\usepackage{threeparttable}
\usepackage{cleveref}

\begin{document}

\title{Probing the impact of radio-mode feedback on the properties of the cool circumgalactic medium}



\author[0000-0002-0196-3496]{Yu-Ling Chang}
\affiliation{Graduate Institute of Astrophysics, National Taiwan University, No. 1, Sec. 4, Roosevelt Road, Taipei 10617, Taiwan}

\author[0000-0001-8857-7020]{Ting-Wen Lan}
\affiliation{Graduate Institute of Astrophysics, National Taiwan University, No. 1, Sec. 4, Roosevelt Road, Taipei 10617, Taiwan}
\affiliation{Department of Physics, National Taiwan University, No. 1, Sec. 4, Roosevelt Rd., Taipei 10617, Taiwan}

\author[0000-0002-7738-6875]{J. Xavier Prochaska}
\affiliation{University of California, Santa Cruz, 1156 High Street, Santa Cruz, CA 95064, USA}
\affiliation{Kavli Institute for the Physics and Mathematics of the Universe, 5-1-5 Kashiwanoha, Kashiwa, 277-8583, Japan}

\author[0000-0002-5166-8671]{Lucas Napolitano}
\affiliation{Department of Physics \& Astronomy, University of Wyoming, 1000 E. University, Dept.~3905, Laramie, WY 82071, USA}

\author[0000-0003-2923-1585]{Abhijeet Anand}
\affiliation{Lawrence Berkeley National Laboratory, 1 Cyclotron Road, Berkeley, CA 94720, USA}

\author{J.~Aguilar}
\affiliation{Lawrence Berkeley National Laboratory, 1 Cyclotron Road, Berkeley, CA 94720, USA}

\author[0000-0001-6098-7247]{S.~Ahlen}
\affiliation{Physics Dept., Boston University, 590 Commonwealth Avenue, Boston, MA 02215, USA}

\author{D.~Brooks}
\affiliation{Department of Physics \& Astronomy, University College London, Gower Street, London, WC1E 6BT, UK}

\author{T.~Claybaugh}
\affiliation{Lawrence Berkeley National Laboratory, 1 Cyclotron Road, Berkeley, CA 94720, USA}

\author[0000-0002-1769-1640]{A.~de la Macorra}
\affiliation{Instituto de F\'{\i}sica, Universidad Nacional Aut\'{o}noma de M\'{e}xico,  Cd. de M\'{e}xico  C.P. 04510,  M\'{e}xico}

\author[0000-0002-4928-4003]{Arjun~Dey}
\affiliation{NSF NOIRLab, 950 N. Cherry Ave., Tucson, AZ 85719, USA}

\author{P.~Doel}
\affiliation{Department of Physics \& Astronomy, University College London, Gower Street, London, WC1E 6BT, UK}

\author[0000-0003-3142-233X]{S.~Gontcho A Gontcho}
\affiliation{Lawrence Berkeley National Laboratory, 1 Cyclotron Road, Berkeley, CA 94720, USA}

\author[0000-0001-9822-6793]{J.~Guy}
\affiliation{Lawrence Berkeley National Laboratory, 1 Cyclotron Road, Berkeley, CA 94720, USA}

\author[0000-0002-0000-2394]{S.~Juneau}
\affiliation{NSF NOIRLab, 950 N. Cherry Ave., Tucson, AZ 85719, USA}

\author[0000-0003-3510-7134]{T.~Kisner}
\affiliation{Lawrence Berkeley National Laboratory, 1 Cyclotron Road, Berkeley, CA 94720, USA}

\author{A.~Lambert}
\affiliation{Lawrence Berkeley National Laboratory, 1 Cyclotron Road, Berkeley, CA 94720, USA}

\author[0000-0003-1838-8528]{M.~Landriau}
\affiliation{Lawrence Berkeley National Laboratory, 1 Cyclotron Road, Berkeley, CA 94720, USA}

\author[0000-0001-7178-8868]{L.~Le~Guillou}
\affiliation{Sorbonne Universit\'{e}, CNRS/IN2P3, Laboratoire de Physique Nucl\'{e}aire et de Hautes Energies (LPNHE), FR-75005 Paris, France}

\author[0000-0003-4962-8934]{M.~Manera}
\affiliation{Departament de F\'{i}sica, Serra H\'{u}nter, Universitat Aut\`{o}noma de Barcelona, 08193 Bellaterra (Barcelona), Spain}
\affiliation{Institut de F\'{i}sica d’Altes Energies (IFAE), The Barcelona Institute of Science and Technology, Campus UAB, 08193 Bellaterra Barcelona, Spain}

\author[0000-0002-4279-4182]{P.~Martini}
\affiliation{Center for Cosmology and AstroParticle Physics, The Ohio State University, 191 West Woodruff Avenue, Columbus, OH 43210, USA}
\affiliation{Department of Astronomy, The Ohio State University, 4055 McPherson Laboratory, 140 W 18th Avenue, Columbus, OH 43210, USA}
\affiliation{The Ohio State University, Columbus, 43210 OH, USA}

\author[0000-0002-1125-7384]{A.~Meisner}
\affiliation{NSF NOIRLab, 950 N. Cherry Ave., Tucson, AZ 85719, USA}

\author{R.~Miquel}
\affiliation{Instituci\'{o} Catalana de Recerca i Estudis Avan\c{c}ats, Passeig de Llu\'{\i}s Companys, 23, 08010 Barcelona, Spain}
\affiliation{Institut de F\'{i}sica d’Altes Energies (IFAE), The Barcelona Institute of Science and Technology, Campus UAB, 08193 Bellaterra Barcelona, Spain}

\author[0000-0002-2733-4559]{J.~Moustakas}
\affiliation{Department of Physics and Astronomy, Siena College, 515 Loudon Road, Loudonville, NY 12211, USA}

\author{A.~D.~Myers}
\affiliation{Department of Physics \& Astronomy, University of Wyoming, 1000 E. University, Dept.~3905, Laramie, WY 82071, USA}

\author[0000-0001-6590-8122]{J.~Nie}
\affiliation{National Astronomical Observatories, Chinese Academy of Sciences, A20 Datun Rd., Chaoyang District, Beijing, 100012, P.R. China}

\author{C.~Poppett}
\affiliation{Lawrence Berkeley National Laboratory, 1 Cyclotron Road, Berkeley, CA 94720, USA}
\affiliation{Space Sciences Laboratory, University of California, Berkeley, 7 Gauss Way, Berkeley, CA  94720, USA}
\affiliation{University of California, Berkeley, 110 Sproul Hall \#5800 Berkeley, CA 94720, USA}

\author[0000-0001-5589-7116]{M.~Rezaie}
\affiliation{Department of Physics, Kansas State University, 116 Cardwell Hall, Manhattan, KS 66506, USA}

\author{G.~Rossi}
\affiliation{Department of Physics and Astronomy, Sejong University, Seoul, 143-747, Korea}

\author[0000-0002-9646-8198]{E.~Sanchez}
\affiliation{CIEMAT, Avenida Complutense 40, E-28040 Madrid, Spain}

\author{M.~Schubnell}
\affiliation{Department of Physics, University of Michigan, Ann Arbor, MI 48109, USA}
\affiliation{University of Michigan, Ann Arbor, MI 48109, USA}

\author[0000-0002-6588-3508]{H.~Seo}
\affiliation{Department of Physics \& Astronomy, Ohio University, Athens, OH 45701, USA}

\author{D.~Sprayberry}
\affiliation{NSF NOIRLab, 950 N. Cherry Ave., Tucson, AZ 85719, USA}

\author[0000-0003-1704-0781]{G.~Tarl\'{e}}
\affiliation{University of Michigan, Ann Arbor, MI 48109, USA}

\author{B.~A.~Weaver}
\affiliation{NSF NOIRLab, 950 N. Cherry Ave., Tucson, AZ 85719, USA}

\author[0000-0002-6684-3997]{H.~Zou}
\affiliation{National Astronomical Observatories, Chinese Academy of Sciences, A20 Datun Rd., Chaoyang District, Beijing, 100012, P.R. China}

\begin{abstract}
We explore the influence of radio-mode feedback on the properties of the cool circumgalactic medium (CGM). To this end, we assemble a statistical sample of approximately 30,000 radio galaxies with background quasars by combining optical spectroscopic measurements of luminous red galaxies (LRGs) and quasars from the year 1 dataset of Dark Energy Spectroscopic Instrument (DESI) and radio sources from the LOw-Frequency ARray Two-metre Sky Survey (LoTSS) DR2 catalog and the Very Large Array Sky Survey (VLASS) quick look catalog. Galaxies with similar optical properties but with no radio counterparts in LoTSS and VLASS are selected as the control group. 
We measure the cool CGM properties of radio galaxies and their control samples traced by MgII absorption lines, including covering fraction, rest equivalent width, and gas kinematics. Our results show no significant difference in the properties of gas around radio galaxies and their control sample, indicating that the operating radio-mode feedback of massive galaxies does not produce detectable effects on the properties of the cool CGM. 
Finally, we show that the CGM of radio galaxies contain a non-negligible amount of cool gas with approximately $10^{10} \rm \, M_{\odot}$. This abundance can place a stringent constraint on the radio-mode feedback models.

\end{abstract}



\section{Introduction} \label{intro}

Feedback from supermassive black holes has been considered as a crucial mechanism driving galaxy evolution \citep[e.g.,][for a review]{Fabian2012, ARAA_simulations}. It is required to be included in simulations in order to explain salient observed properties of galaxies, especially the massive end of galaxy stellar mass function \citep[e.g.,][]{Benson2003,Bower2006, Croton2006, Hirschmann2014, Beckmann2017, Kondapally2023}. It is also a key component to reconcile the so-called cooling flow problem \citep[e.g.,][]{Cowie1977, Fabian1977}, indicating that the gas in massive halos is expected to cool in a short period based on the X-ray emission of hot gas and to condense into new stars — a trend which is not observed. The energy output from the feedback mechanisms has been treated as the main source to balance this gas cooling \citep{Rafferty2006, Nulsen2007, Hlavacek-Larrondo2012, Fabian2012, McNamara2012}. In other words, in the current galaxy formation theory, feedback from supermassive black holes regulates the amount of gas in and out of galaxies as well as maintains the thermal content of the circumgalactic medium (CGM) \citep[e.g.,][for a review]{Tumlinson2017}. 

One line of evidence showing the impact of feedback on the CGM properties is from observations of galaxy clusters, such as Perseus and Cygnus A, in X-ray and radio wavelengths \citep[e.g.,][]{Boehringer1993, Carilli1994,Fabian2003, McNamara2005, Birzan2008, Gitti2012, Pandge2019, Eckert2021}. These observations demonstrate that hot gas ($T\sim10^{6}-10^7$ K) distribution in these galaxy clusters is clearly affected by the radio jets from supermassive black holes, the so-called radio-mode feedback \citep[e.g.,][for a review]{McNamara2012, Heckman2014, Hardcastle2020}. Based on the observations, the estimated required energy to create the cavity of hot gas and the energy output from the radio-mode feedback are consistent with each other. This indicates that radio-mode feedback from supermassive black holes can prevent the cooling of the hot atmosphere, regulate the gas supply, and indeed account for the quenching of massive galaxies. Recent galaxy simulations and subgrid models incorporating radio-mode feedback have successfully reproduced the observed luminosity function of massive galaxies \citep[e.g.,][]{Bower2006, Croton2006, Gabor2011, Dave2012, TNGradiofeedback}. 

While the influence of radio-mode feedback on the properties of the CGM has been studied, the observations have mostly focused on (1) massive systems, such as galaxy clusters ($M_{halo}>10^{14} M_{\odot}$), and (2) the hot gas ($T\geq10^{6}$ K) properties via X-ray observations \citep[e.g.,][]{McNamara2007}. There are limited observational efforts probing relatively less massive halos and the properties of the cool CGM ($T\sim10^{4}$ K), which can also be sensitive to radio-mode feedback and be a strong constraint on the corresponding models \citep{Huang2016, Smailagic2023}. 

To probe the cool CGM properties of galaxies with radio-mode feedback in operation, one can utilize absorption line spectroscopy \citep[e.g.,][]{Tumlinson2017}. That is, by constructing a sample of galaxies with radio emission having bright sources, such as quasars, in the background with lines of sight intercepting the CGM of the galaxies. However, the number of galaxies with radio emission and with background quasars detectable by wide-field radio surveys is rare. 
This limits the observational ability to probe the impact of radio-mode feedback on the properties of the cool CGM, and only a limited number of relevant studies \citep[e.g.,][]{Kauffmann2017} to date.

In this work, we overcome this observational limitation by utilizing the large optical spectroscopic dataset obtained by the Dark Energy Spectroscopic Instrument (DESI) survey \citep{DESI2016a, DESI_instr} and the large source catalogs provided by two radio sky surveys, the LOFAR Two-metre Sky Survey \citep[LoTSS,][]{LOTSS} and the Very Large Array Sky Survey \citep[VLASS,][]{VLASS}.
Through this approach, we study the properties of the cool CGM around massive galaxies, such as luminous red galaxies (LRGs), with radio-mode feedback in action and compare the gas properties with that of a control sample without radio emission. This adds a new parameter which has not been extensively explored in previous studies of the properties of gas around LRGs \citep[e.g.,][]{Zhu2014, Huang2016, Lan2018, Anand2021}.

Combining radio emission signals with optical galaxy properties, we construct a sample of radio galaxies with $z>0.4$ with background quasars as well as a control sample without radio emission and measure the properties of the cool CGM traced by MgII absorption lines $\rm \lambda\lambda~2796, 2803 \AA$, an absorption line species which has been widely used to probe the CGM content across cosmic time \citep[e.g.,][]{Lanzetta1987, Nestor2005, Nielsen2013, Zhu2013b, Lan14, Raghunathan2016, Huang2016, Chen2017, Lan2018, Lan2020, Anand2021, Zou2021, Anand2021, Anand2022, Zou2024}, and explore the correlation between the gas properties and the presence of radio emission, a proxy of radio-mode feedback in action.

This paper is organized in the following way: Section~\ref{data} describes our sample and data, and Section~\ref{method} presents how we obtain the properties of MgII absorption lines. 
We show our results in Section~\ref{results} and discuss the implication in Section~\ref{discussion}. 
Section~\ref{conclusion} summarizes the paper. Throughout the paper, we adopt a Flat-$\Lambda$CDM cosmology with the following parameters: $\Omega_{\rm m}=0.3~{\rm and}~H_{0}=70~{\rm km}~{\rm s}^{-1}~{\rm Mpc}^{-1}$. 

\begin{figure*}
\begin{center}
\includegraphics[width=0.99\linewidth]{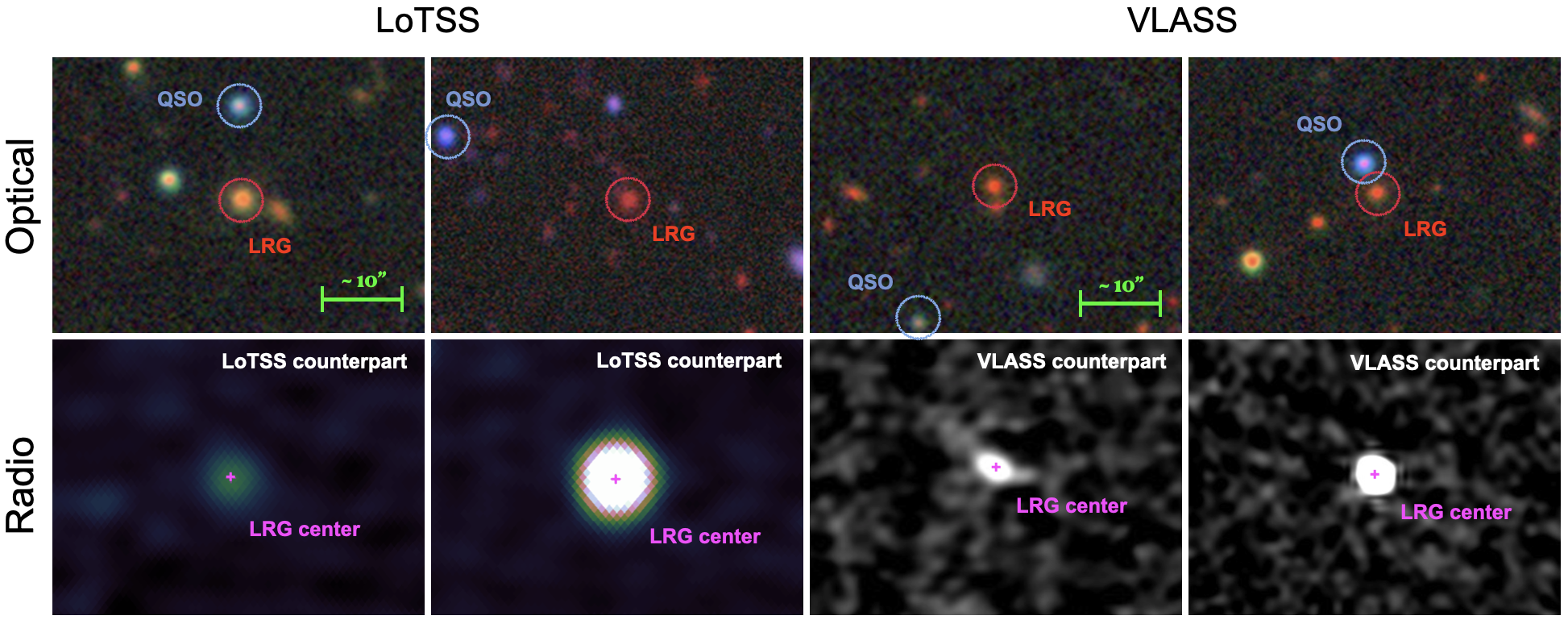}
\end{center}
\caption{Examples of LRG-QSO pairs with radio counterparts in LoTSS survey (left) and VLASS (right).The upper panels are the optical images from the DESI Legacy Imaging Surveys \citep{Dey2019} and the lower panels are the radio images from the two radio surveys.}
\label{pair_radio}
\end{figure*}

\begin{figure*}
\begin{center}
\includegraphics[width=0.99\linewidth]{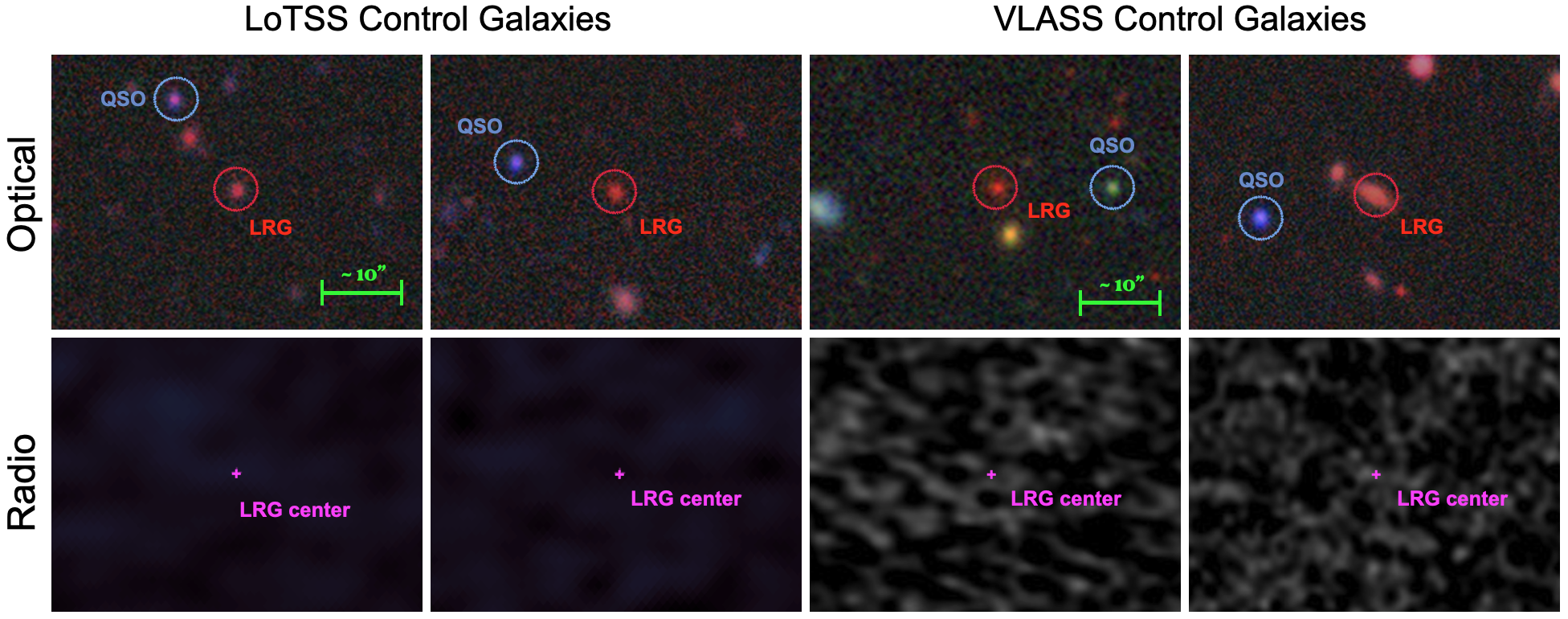}
\end{center}
\caption{Examples of LRG-QSO pairs without radio counterparts in LoTSS survey (left) and VLASS (right), i.e. the control samples. The upper panels are the optical images from the DESI Legacy Imaging Surveys \citep{Dey2019} and the lower panels are the radio images from the two radio surveys.}
\label{pair_control}
\end{figure*}

\begin{figure*}
\begin{center}
\includegraphics[width=0.9\linewidth]{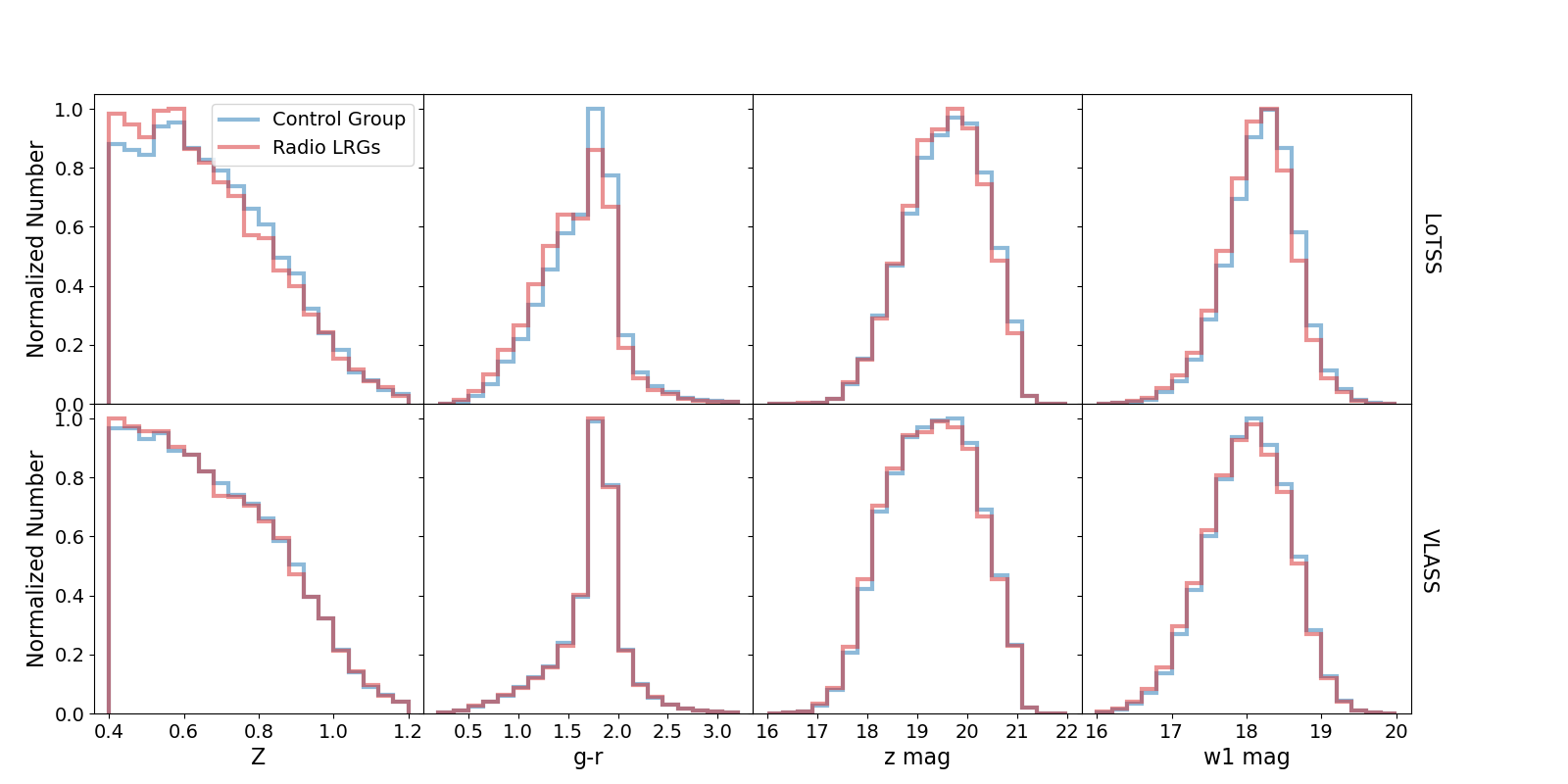}
\end{center}
\caption{Distributions of redshift, color, z band magnitude, and w1 band magnitude (from left to right) for the experimental group and the control group from the LoTSS catalog and the VLASS catalog. The upper panel shows the properties of the LRGs from the LoTSS samples, while the lower panel shows those of the LRGs from the VLASS. The red and blue lines represent the radio groups and the control groups, respectively. }
\label{control_dist}
\end{figure*}

\section{Data} \label{data}

\subsection{DESI} \label{desi}

The Dark Energy Spectroscopic Instrument (DESI) is mounted on the Mayall 4-meter telescope at Kitt Peak National Observatory \citep{DESI2016a,DESI2016b}. It consists of 5000 fibers \citep{DESI_focul_plane} with spectral wavelength coverage ranging from $3600 - 9800{\rm \AA}$ and a resolution of 2000-5000 \citep{DESI_instr,DESI_pipeline}. The spectroscopic classifications and redshifts for each DESI source are determined using the Redrock pipeline\footnote{\url{https://github.com/desihub/redrock}} \citep[Bailey et al, in preparation,][]{Brodzeller2023}. 
The survey includes dark-time and bright-time observations for different type of sources \citep{DESI_survy_ops}. 
The bright time observations cover stars in our Milky Way \citep{DESI_MWS} and bright galaxies \citep[BGS,][]{DESI_BGS,DESI_BGS_AGN}, while the dark time observations cover relative faint extragalactic sources, including luminous red galaxies \citep[LRGs,][]{DESI_LRG}, emission line galaxies \citep[ELGs,][]{DESI_ELG}, and quasars \citep{DESI_QSO}. 
 
 The target selection scheme for the DESI sources \citep{DESI_target} and the corresponding redshift performance with the DESI observations have been tested and validated during the survey validation (SV) phase, including observations with long exposure times  (Survey Validation 1, SV1) and observations covering about one percent of the whole DESI survey (One-percent survey) \citep{DESI_SV, DESI_galaxy_VI,DESI_quasar_VI}. In this work, we use the LRG and quasar data from SV \citep[Early Data release,][]{DESI_EDR_cat} and the first year of the main survey. The cosmological constraints from the baryon acoustic oscillations measurements with the DESI first year main survey data have been published \citep{DESI2024III,DESI2024IV,DESI2024VI}.

\textbf{The DESI LRG survey:}
LRGs are massive passive galaxies selected based on their colors and brightness with redshift range $0.3 < z \leq 1.1$ \citep{DESI_LRG}.  
For our analysis, we select LRGs without any warning flag from the DESI pipeline, {\tt ZWARN} = 0\footnote{\url{https://github.com/desihub/redrock/blob/main/py/redrock/zwarning.py}} and with {\tt DELTACHI2} $\Delta \chi ^2> 15$, a difference of $\chi ^2$ values the between the second and first best-fit models, to ensure a robust redshift estimation \citep{DESI_LRG,DESI_galaxy_VI}. 
We further select LRGs with $0.4<z<1.2$ to ensure that MgII absorption lines are within the DESI wavelength coverage. 
The above selection leads to a sample of  $\sim2.2\times10^{6}$ LRGs. 

\textbf{The DESI QSO survey:}
The DESI survey observes quasars selected with the random forest algorithm across a wide redshift range from redshift 0.5 to beyond redshift 2 \citep{DESI_QSO}. Redshift $>2.1$ quasars are primarily used as background light for probing the baryonic acoustic oscillation signals via detecting Lyman alpha forest \citep[e.g.,][]{DESI_Lya}. Low redshift quasars can be used as direct tracers of matter distribution. Similarly to the LRG selection, we apply the criteria {\tt ZWARN} = 0 and {\tt DELTACHI2} $> 15$ to select DESI QSOs, ensuring that we include only quasars with reliable redshift measurements from the pipeline \citep{DESI_QSO,DESI_quasar_VI}. 
We also exclude quasars with spectral signal-to-noise lower than 1, a spectral quality too low to detect absorption lines. Our quasar sample includes $\sim1.2\times10^{6}$ sources in total. 

\subsection{Radio catalogs} \label{radio_cat}
\textbf{The LOFAR Two-meter Sky Survey:}
The LOFAR Two-meter Sky Survey \citep[LoTSS,][]{LOTSS} is an ongoing radio wavelength (120-168 MHz) sky survey, aiming to cover the entire northern sky. The angular resolution is about $6 \arcsec$ with sensitivity down to  $\sim 70-100$~mJy. In this work, we make use of the second data release of LoTSS, LoTSS-DR2 \citep{LOFAR-DR2}, which covers approximately 27\% of the northern sky with $\sim 4\times10^{6}$ radio sources. 
LoTSS-DR2 sky coverage overlaps with the sky coverage of the DESI survey in the northern sky, making the DR2 catalog a suitable resource for identifying radio counterparts of the DESI galaxies.

\textbf{VLASS Quick Look:}
The Karl G. Jansky Very Large Array Sky Survey \citep[VLASS,][]{VLASS} aims to observe the entire northern sky of $\delta > -40^\circ$ with three epochs at $2~{\rm GHz}<\nu<4~{\rm GHz}$ frequency band with $2.5\arcsec$ angular resolution. We use the VLASS data based on the quick look (QL) imaging \citep{VLASS_QL} from the first two epochs at 3~GHz\footnote{\url{https://cirada.ca/vlasscatalogueql0}}, including $\sim 3.5\times10^6$ sources. The first two epochs of VLASS QL data exhibit a sensitivity comparable to that of FIRST, $\sim 1$~mJy, and cover the entire VLASS footprint. 

\section{Analysis} 
\label{method}
\subsection{LRG-QSO Pairs} 
To study the CGM of radio galaxies, we construct a sample of radio-detected DESI LRGs with background quasars. To this end, we utilize the radio-optical crossmatch catalog for the LoTSS DR2 \citep{Hardcastle2023} and match those optical identifications and DESI LRGs with a $0.75\arcsec$ search radius \citep{Rosario2020}. Additionally, we cross-match the LRGs with radio sources in the VLASS quick-look catalogs using a 2.5\arcsec matching radius. 
Around $14\%$ and $1.6\%$ of the DESI LRGs have detected radio counterparts in LoTSS and VLASS, respectively. We note that the different detection fractions of LoTSS and VLASS counterparts are due to the sensitivities of the two surveys\footnote{LoTSS is $\sim 5$ times deeper than the VLASS when converting flux at 3~GHz, assuming $\alpha =0.5$. We have used the radio flux-limited samples (1~mJy for LoTSS and 3~mJy for VLASS) to perform the analyses, which yield consistent results with the whole samples.}. The impact parameter (projected distance, $r_p$) of the galaxies and background quasar sightlines are within 1000 kpc. To avoid potential contamination of radio emissions from background quasars, we only include systems with the distances between quasars and the centers of radio emission being two times larger than the distances between LRGs and the centers of radio emission.
That is, $d_{qso} \geq 2\times d_{lrg}$, where $d_{lrg}$ is the angular separation between the LRG and the radio source associated with it and $d_{qso}$ is the separation between the QSO and the associated radio source for the foreground LRG. 

In order to isolate the effects of radio-mode feedback on the CGM properties of radio galaxies, we create a set of control galaxies for LoTSS pairs and VLASS pairs. 
For each DESI radio LRG in the LoTSS and VLASS catalogs, two and four control LRGs are respectively selected based on the combined nearest Euclidean distances of four optical properties: $z$-band flux, WISE W1 flux, $g-r$ color, and redshift\footnote{Because of the limited number of sources in the LoTSS footprint available for selection as the control group, which is only one-sixth the size of the VLASS control group, adding more control galaxies for each LoTSS LRG would lead to a mismatch in the distribution of optical properties compared to the radio sample.}. All the control galaxies are required to have no radio detection from LoTSS or VLASS within $10\arcsec$. For the LoTSS control sample, we select DESI LRGs located in the sky coverage of the LoTSS DR2 footprint without detected radio emission, while for the VLASS control sample, we select DESI LRGs without detected radio emission in the DESI dataset given that the VLASS survey covers most of the DESI footprint. For the final control samples, we remove duplicate control systems selected more than one time from different radio-detected LRGs. 

Figure~\ref{pair_radio} and \ref{pair_control} show examples of LRG-QSO pairs and their corresponding radio counterparts for both the experimental groups and the control groups.
Figure~\ref{control_dist} shows the observed properties of the radio LRGs and the control samples, demonstrating the two groups have similar optical properties. 
The main difference is that the control galaxies do not have radio emission detected in the corresponding surveys.
We check the radio morphology for the radio-detected LRGs. For the LoTSS-detected sources, we utilize Equation 2 from \citet{LOFAR-DR2} to identify extended sources. For VLASS-detected sources with peak flux$>$ 3mJy/beam, we find the extended source using the deconvoluted-size criteria outlined in \citet{VLASS_QL}. Our final samples include $\sim 8\%$ and $\sim 21\%$ of extended LRGs for LoTSS and VLASS, consistent with the fraction for all the sources in both radio catalogs.

\begin{figure} [h!]
\begin{center}
\includegraphics[width=0.8\linewidth]{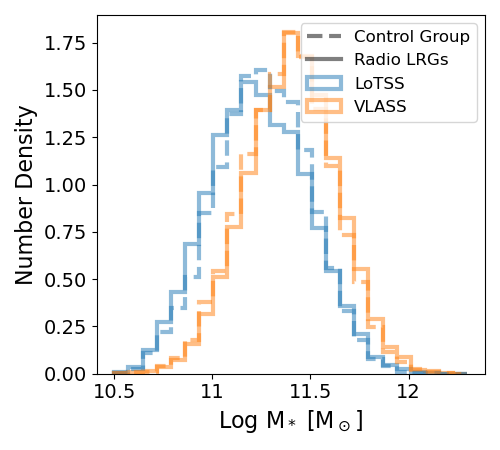}   
\end{center}
\caption{Stellar mass distribution of our samples from CIGALE SED fitting. The red lines show the distribution of the LRGs from the LoTSS samples, while the blue lines show that of the LRGs form the VLASS samples. The dashed and solid lines represent the control groups and radio groups, respectively.}
\label{stellar}
\end{figure}

We also estimate the stellar masses of the LRGs by performing spectral energy distribution fitting with $g$, $r$, $z$, and W1 fluxes using the CIGALE \citep{CIGALE}. To model the spectra and obtain the stellar properties with the CIGALE, we use \citet{BC2003} model and assume a \citet{Chabrier2003} initial mass function. 
The stellar mass distributions are shown in Figure~\ref{stellar}. The results show that stellar mass distributions between the radio LRGs and their control galaxies are consistent with each other. We note that the VLASS sample has a higher median stellar mass ($\sim 0.2$ dex) 
than the LoTSS sample, which is due to the difference of the sensitivity between the two surveys. The VLASS, with its shallower depth, primarily detects brighter radio galaxies which tend to be more massive\citep{Jarvis2002, Best2005}.



We further include the following selections to obtain clean spectral regions for detecting MgII absorption lines associated with the LRGs:
\begin{itemize}
    \item To avoid the possible contamination of QSO-associated absorbers, the velocity difference between LRGs and quasars is required to be $\delta v \geq 6,000~{\rm km/s}$ \citep{DESI_EDR_MgII}.
    \item We exclude the spectral regions close to the CIV emission lines of quasars with $\delta v \lesssim 10000 {\rm km/s}$ to avoid broad CIV absorption features.
    \item We also exclude pairs with high redshift quasars with the wavelength region of MgII absorption lines at the galaxy redshifts overlapping with the Lyman alpha forests with $\delta v \lesssim 12000 {\rm km/s}$. 
\end{itemize}

\begin{table} [h!]
\begin{threeparttable}[b]
\caption{Number of pairs with all the selection cuts.}
\label{tab_pairs}
    \begin{tabular}{lcc}
    \hline
    \multicolumn{3}{c}{QSO-LRG pairs within $r_p=1000$~kpc: 1279006} \\    
      & LoTSS & VLASS \\ 
    \hline
    \hline
    \multicolumn{3}{c}{Select radio-detected and the control group} \\    
    radio LRGs&  35720 & 21014 \\
    control group& 48923\tnote{a} & 77325\tnote{b} \\
    \hline
    \multicolumn{3}{c}{After excluding QSO-associated absorbers} \\    
    radio LRGs&  35197 & 20629 \\
    control group& 48165 & 76025 \\
    \hline
    \multicolumn{3}{c}{After excluding region close to CIV emission lines} \\    
    radio LRGs&  31585 & 18601 \\
    control group& 43301 & 68343 \\
    \hline
    \multicolumn{3}{c}{After excluding region within Ly$\alpha$ forest} \\    
    radio LRGs&  26949 & 16013 \\
    control group& 37379 & 59017 \\
    \hline
    \end{tabular}
\begin{tablenotes}
\item [a]Two control galaxies with the closest optical properties to each LoTSS-detected LRG are selected. LoTSS covers a much smaller area of the sky compared to VLASS, leading to a limited number of unique control galaxies selected.
\item [b] Four control galaxies with the closest optical properties to each VLASS-detected LRG are selected. 
\end{tablenotes}    
\end{threeparttable}
\end{table}

Table~\ref{tab_pairs} summarizes the number of pairs after each selection criterion. 
The final sample consists of 26949 and 16013 radio LRG-quasar pairs from LoTSS and VLASS respectively and 37379 and 59017 LRG-quasar pairs as the control sample. 
We note that 9802 pairs overlap between the LoTSS-selected and VLASS-selected samples. 

\subsection{MgII absorbers} \label{absorbers}
To detect absorption lines in quasar spectra, the first step is to estimate the continuum intrinsic to the source. We follow the procedure applied in \citet{Zhu2013b}, utilizing the Non-negative Matrix Factorization (NMF) \citep{Lee1999,Zhu2016} method to estimate the quasar spectral energy distribution with twelve NMF eigenspectra bases obtained in \citet{Zhu2013b}. We further apply a median filter with 85 pixels \citep{DESI_EDR_MgII} to remove the small-scale fluctuation on the normalized spectra. 

With the normalized quasar spectra, we develop an automatic algorithm to detect MgII absorption lines around LRGs with the following procedure:
\begin{enumerate}
    \item We use a matched filter method with a MgII doublet profile based on SDSS composite spectra \citep{Lan17} and convolve the absorption line profile with the pixel values of each normalized spectrum across velocity window between -1000 km/s and 1000 km/s around the LRG rest-frame.
    We also apply this calculation to the error arrays to estimate the uncertainty of the convolved spectra. 
    \item From the convolved spectra and the corresponding uncertainty, we identify spectra with pixels having maximum S/N greater than 2 and consider the velocity of the pixel with maximum S/N as the central velocity of the MgII absorber candidate. 
    
    \item We use a double Gaussian profile to fit the original normalized spectra at the central velocities of MgII absorber candidates and measure the absorption line properties, including MgII rest equivalent widths $W_{0, \lambda 2796, 2803}$, line widths, and the final best-fit central velocities of the MgII absorber candidates as well as their corresponding uncertainties. 

    \item Finally, we consider MgII absorber candidates with S/N of $W_{0,\lambda2796}\geq 3$ and S/N of $W_{0,\lambda2803}\geq 2$ as detected MgII absorption lines. 
\end{enumerate}
Other absorption line species, such as FeII and CIV, can accidentally fall within the search window and mimic the MgII absorption line signals. To remove such possible contamination, we use the MgII line ratio ($R=W_{0,\lambda2796}/W_{0,\lambda2803}$) as the criterion. More specifically, we calculate the line ratio distribution as a function of $W_{0,\lambda2796}$, using the DESI EDR MgII absorber catalog \citep{DESI_EDR_MgII} as a reference. For a given $W_{0,\lambda2796}$, we remove MgII absorbers with the line ratio values being outside of 0.135-99.865 percentile of the $R$ distribution from DESI EDR MgII catalog. This removes approximately $3\%$ of the systems. 

Our final LRG MgII absorber samples include 957 and 1318 absorbers selected from LoTSS and VLASS catalogs, respectively.
Approximately 400 and 300 of these absorbers are associated with radio emissions from the LoTSS and VLASS. 
The numbers are summarized in Table~\ref{tab:absorbers}. 

\begin{figure*}
\begin{center}
\includegraphics[width=0.85\linewidth]{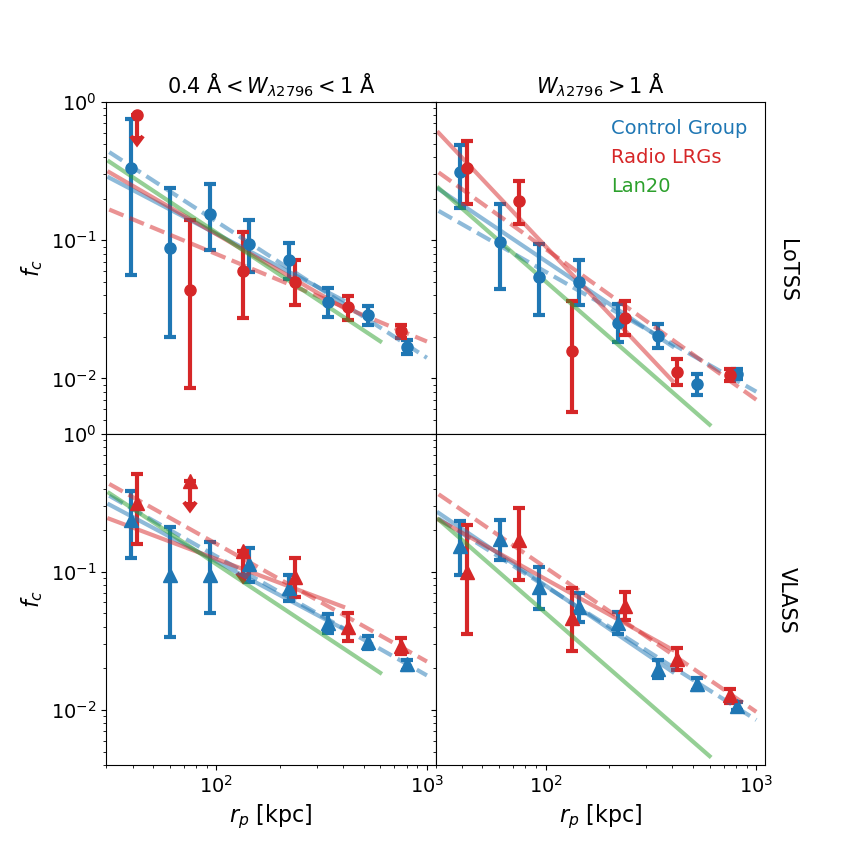}
\end{center}
\caption{Covering fraction ($f_c$) as a function of impact parameter ($r_p$). The upper panels show the $f_c$ for LRGs from the LoTSS samples, while the lower panels show that for LRGs from the VLASS samples. The $f_c$ of weak and strong absorbers are depicted correspondingly in the left and right panels. The red and blue points represent the radio groups and control groups, respectively. The solid lines and dashed lines illustrate the best-fit results for $r_p \leq 400~{\rm kpc}$ and $r_p \leq 1000~{\rm kpc}$. Green lines show the $f_c$ of passive galaxies in \citet{Lan2020}. The upper limits are determined at a 3-sigma confidence level, and the errors are estimated based on binomial statistics \citep{Gehrels1986}.}
\label{cov_rp}
\end{figure*}

\begin{table} [h!]
\begin{center}
\caption{The number of pairs used to detect MgII absorption lines and the identified absorbers in both experimental and control groups from each radio catalog.}
    \begin{tabular}{lcccc}
    \hline
    \hline
      & \multicolumn{2}{c}{Absorbers} & \multicolumn{2}{c}{LRG-QSO pairs} \\ 
      & LoTSS & VLASS & LoTSS & VLASS \\
    \hline
    Radio LRGs&  415 & 305  & 26949  & 16013 \\
     & $1.5\%$ & $1.9\%$ & $\cdots$ & $\cdots$ \\
    \hline
    Control Group& 542 & 1013 & 37379 & 59017 \\
    & $1.5\%$ & $1.7\%$ & $\cdots$ & $\cdots$ \\
    \hline
    \end{tabular}
\label{tab:absorbers}
\end{center}
\end{table}

\begin{table} [h!]
\begin{center}
\caption{Number of detected MgII absorption lines with different spectral S/N cuts. The S/N cuts for absorbers with $0.4 \leq W_{0,\lambda2796} < 1~{\rm \AA}$ and absorbers with $W_{0,\lambda2796} \geq 1~{\rm \AA}$ are 6 and 2.5, respectively. Here we also show the number of all detected absorbers with a uniform S/N cut of 6.}
    \begin{tabular}{lcccc}
    \hline
    \hline
     & \multicolumn{2}{c}{S/N cuts by $W_{0,\lambda2796}$} & \multicolumn{2}{c}{All S/N~$\geq 6$}\\  
      & LoTSS & VLASS & LoTSS & VLASS \\
    \hline
    Radio Pairs&  264 & 215  &  234 & 177  \\
    Control Sample& 364 & 785 & 322 & 591 \\  
    \hline
    \end{tabular}
\label{tab:sncut_absorbers}
\end{center}
\end{table}

\begin{table*} [h!]
\begin{center}
\caption{Best-fit parameters for covering fraction $f_c$.}
    \begin{tabular}{lcccc}
    \hline
    \multicolumn{5}{c}{\tiny} \\
    & \multicolumn{4}{c}{\large $r_p \leq 400~{\rm kpc}$} \\
    & \multicolumn{2}{c}{\large LoTSS} & \multicolumn{2}{c}{\large VLASS} \\      
    \hline
    \hline
    & \multicolumn{4}{c}{$0.4 \leq W_{0,\lambda2796} < 1~{\rm \AA}$} \\    
      & $\gamma$ & ${\rm A}$  & $\gamma$ & ${\rm A}$ \\   
    Radio LRGs&  $-0.88\pm0.74$ & $0.11\pm0.11$ & -$0.57\pm0.43$ & $0.12\pm0.07$  \\
    Control Group& $-0.80\pm0.39$ & $0.11\pm0.05$ & $-0.81\pm0.24$ & $0.12 \pm0.03$ \\
    \hline
        & \multicolumn{4}{c}{$W_{0,\lambda2796} \geq 1 ~{\rm \AA}$} \\ 
      & $\gamma$ & ${\rm A}$  & $\gamma$ & ${\rm A}$ \\   
    Radio LRGs&  $-1.62\pm0.27$ & $0.09\pm0.02$ & $-0.85\pm0.34$ & $0.09\pm0.03$  \\
    Control Group& $-1.02\pm0.24$ & $0.07\pm0.02$ & $-1.03\pm0.16$ & $0.08 \pm0.01$ \\
    \hline
    \multicolumn{5}{c}{\tiny} \\
    & \multicolumn{4}{c}{\large $r_p \leq 1000~{\rm kpc}$} \\
    & \multicolumn{2}{c}{\large LoTSS} & \multicolumn{2}{c}{\large VLASS} \\      
    \hline
    \hline   
    & \multicolumn{4}{c}{$0.4 \leq W_{0,\lambda2796} < 1~{\rm \AA}$} \\        
      & $\gamma$ & ${\rm A}$  & $\gamma$ & ${\rm A}$ \\   
    Radio LRGs&  $-0.63\pm0.24$ & $0.08\pm0.04$ & -$0.85\pm0.19$ & $0.16\pm0.07$  \\
    Control Group& $-0.99\pm0.15$ & $0.14\pm0.04$ & $-0.86\pm0.10$ & $0.13 \pm0.03$ \\
    \hline
    & \multicolumn{4}{c}{$W_{0,\lambda2796} \geq 1~{\rm \AA}$} \\    
     & $\gamma$ & ${\rm A}$  & $\gamma$ & ${\rm A}$ \\   
    Radio LRGs&  $-1.09\pm0.12$ & $0.09\pm0.02$ & $-1.04\pm0.16$ & $0.11\pm0.03$  \\
    Control Group & $-0.87\pm0.11$ & $0.06\pm0.01$ & $-0.97\pm0.07$ & $0.08 \pm0.01$ \\
    \hline

    \end{tabular}
\label{cov_fit_para}
\end{center}
\end{table*}

\begin{table*} [h!]
\setlength\tabcolsep{3pt}
\begin{center}
\caption{Best-fit parameters for the number density distribution of $W_{0,\lambda2796}$ for absorbers within $r_p \leq 400~{\rm kpc}$.}
    \begin{tabular}{lcccc}
    \hline
    \hline
    & \multicolumn{2}{c}{LoTSS} & \multicolumn{2}{c}{VLASS} \\      
      & $W^{*}$ & $N^*$  & $W^{*}$ & $N^*$ \\   
    \hline
    Radio LRGs&  $0.60\pm0.19$ & $3.06\pm1.61$ & $0.48\pm0.14$ & $4.78\pm2.64$  \\
    Control Group& $0.56\pm0.10$ & $3.56\pm1.19$ & $0.69\pm0.09$ & $2.49 \pm0.53$ \\
    \hline
    \end{tabular}
\label{rew_para}
\end{center}
\end{table*}

\subsection{Completeness of the sample} \label{complete}
The detectibility of absorption lines depends on the S/N of the quasar spectra and the strengths of the absorption lines \citep[][]{Zhu2013a,Anand2021}. The incompleteness due to undetected absorption lines affects the estimation of intrinsic properties of MgII absorbers, including the incidence rate around galaxies. To assess the effect of incompleteness and correct for it, we perform Monte Carlo simulation by creating mock MgII absorption lines as a function of $W_{0,\lambda2796}$ and S/N of the spectra. The mock MgII absorption profiles are based on a set of composite spectra built by stacking MgII absorption lines detected in DESI year 1 quasar sample (Napolitano et al. in prep.) with the pipeline developed by \citet{DESI_EDR_MgII}. In the simulations, we have also taken the DESI spectral resolution at different redshifts into account. Finally, we estimate the detection rate ($\rm N_{recovered}/N_{input}$) of MgII absorbers as a function of absorber redshift, $W_{0,\lambda2796}$, and S/N of the spectra. The details of the detection rate are shown in the Appendix.
The inverse of the detection rate is used as a weight, $w$,  to correct for the undetected absorbers due to the noise of the spectra.

However, using the weights to recover the none-detection will reach its limitation due to the fact that for spectra with too low S/N, one can not detect any absorbers with $W_{0,\lambda2796}$ below the noise level. To avoid reaching such a limitation, we adopt a conservative selection. For $0.4 \leq W_{0,\lambda2796} < 1 \rm\, \AA$, we only consider the detection from the normalized spectra with $S/N\geq 6$ and for $W_{0,\lambda2796}\geq1 \rm\, \AA$, we only consider the detection from the normalized spectra with $S/N\geq 2.5$. 
These two threshold values are determined based on our simulations showing that with such cuts, using weights to correct for the undetected absorbers is effective. 
Finally, we summarize our samples with the spectral S/N selections in Table~\ref{tab:sncut_absorbers}. 
In the following, we define ``weak" absorbers as having $0.4 \leq W_{0,\lambda2796} < 1 \rm\, \AA$ and ``strong" absorbers as those with $W_{0,\lambda2796} \geq 1 \rm\, \AA$.

\section{Results} \label{results}
\subsection{MgII covering fraction} \label{covering}
We first explore the radial distribution of gas traced by MgII absorption lines around DESI radio LRGs and the control sample. To this end, we measure the MgII covering fraction, $f_c$, indicating the probability of detecting absorbers around galaxies with 
\begin{equation}
    f_{c} = \frac{\sum_{i}^{N_{abs}} w_{i}}{N_{\rm quasars}},
\end{equation}
where $w_{i}$ is the weight for each detected absorber, $N_{abs}$ is the total number of detected absorbers, and $N_{\rm quasar}$ is the total number of LRG-QSO pairs for detecting MgII absorbers. 

\begin{figure*}
\begin{center}
\includegraphics[width=0.85\linewidth]{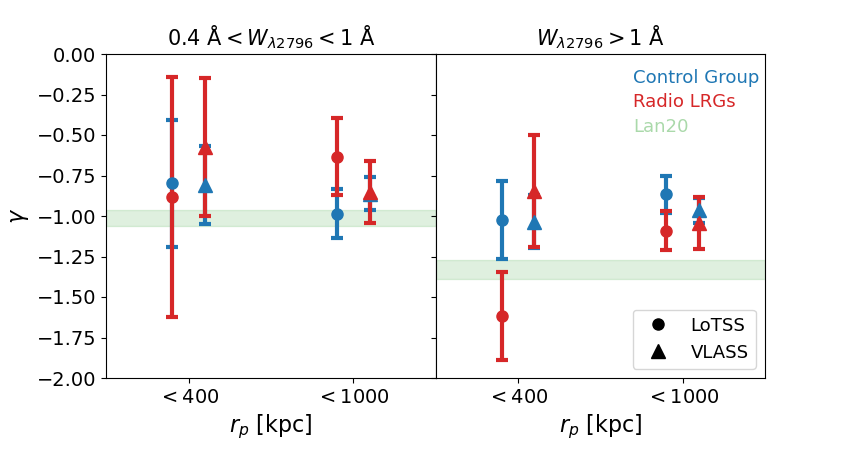}
\end{center}
\caption{Best-fit slopes for covering fraction $f_c$. The left and right panels illustrate outcomes specific to weak and strong absorbers. The circles represent LRGs from the LoTSS samples, and the triangles represent those from the VLASS samples. The blue, red, and green data show the results for the control groups, radio groups, and galaxies from \citet{Lan2020}, respectively.}
\label{cov_para}
\end{figure*}

Figure~\ref{cov_rp} shows the $f_c$ as a function of ${\rm r}_p$ with two $W_{\lambda 2796}$ bins for strong ($W_{0,\lambda2796}\geq 1 \rm \, \AA$) and weak absorbers ($0.4 \leq W_{0,\lambda2796}<1 \rm \, \AA$) with the upper and lower panels indicating the results for DESI LRGs with LoTSS and VLASS respectively. The red and blue data points show the $f_c$ measurements of radio galaxies and the corresponding control sample respectively. 
As shown in Figure~\ref{cov_rp}, the covering fraction of strong and weak absorbers decreases with $r_{p}$ which is consistent with the overall gas distribution traced by MgII absorbers around galaxies \citep{Zhu2014, Huang2016, Lan14, Lan2018, Lan2020, Huang2021, Anand2021}. Moreover, we quantify the possible difference in cumulative distributions of the covering fraction between the radio and the control samples with the Kolmogorov-Smirnov test (KS test) in 2 dimensions \citep{Peacock1983,Fasano1987}\footnote{\url{https://github.com/Gabinou/2DKS}}, $W_{\lambda 2796}$ and $r_p$. The p-values for the weak and strong absorbers for the LoTSS are 0.420 and 0.275, respectively. For the VLASS, the p-value is 0.668 for the weak absorbers and 0.985 for the strong absorbers. These results indicate no detectable difference in the $f_c$ between radio LRGs and the control galaxies for both strong and weak absorbers around radio LRGs and the control sample. 


To summarize the gas distribution, we use a power-law function,
\begin{equation}
    f_{c} = A\times \bigg(\frac{r_{p}}{100~{\rm kpc}}\bigg)^{\gamma},
\end{equation}
to describe the $f_c$ measurements for both $r_p \leq 400~{\rm kpc}$ (solid lines) and the extended range up to $r_p \leq 1000~{\rm kpc}$ (dashed lines), as illustrated in Figure~\ref{cov_para}. 
The best-fit parameters are summarized in Table~\ref{cov_fit_para}. 
Figure~\ref{cov_para} shows the best-fit $\gamma$ values of the covering fraction and the $\gamma$ values obtained in \citet{Lan2020} within 600 kpc, demonstrating that the measurements within virial radius of dark matter halos are consistent. 
The fitting result indicates that the $f_c$ distribution becomes flatter when considering gas distribution at larger scales $r_p \leq 1000~{\rm kpc}$ for both weak and strong absorbers. This behavior can be explained by the transition between one-halo and two-halo terms of the matter distribution \citep[e.g.,][]{Zhu2014}. 


\begin{figure*} 
\begin{center}
\includegraphics[width=0.8\linewidth]{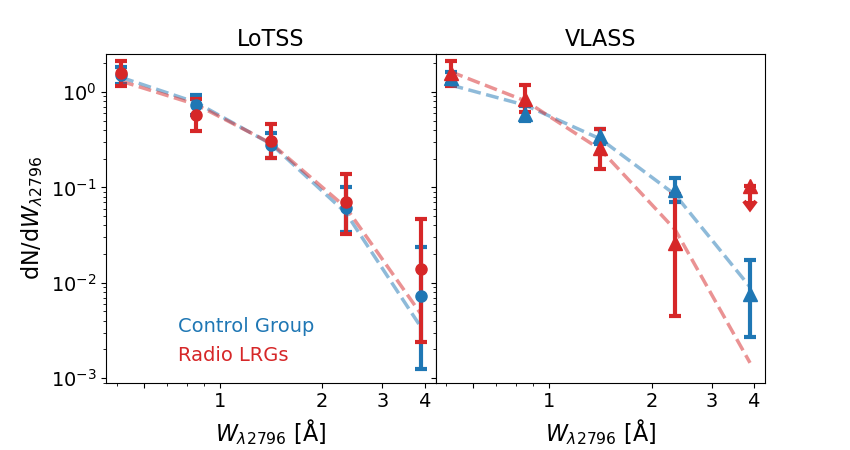}
\end{center}
\caption{Number density distribution of the rest equivalent widths $W_{0,\lambda2796}$ for absorbers with $r_p\leq400$~kpc. The left and right panels show the distributions, respectively corresponding to the absorbers of LRGs from the LoTSS samples and the VLASS samples. The red and blue colors represent the radio groups and the control groups, respectively. The dashed lines are the best-fit exponential functions. Errors are the confidence levels derived from Poison statistics \citep{Gehrels1986}.}
\label{rew_dist}
\end{figure*}

\subsection{Rest equivalent width distributions}
Besides the covering fraction, we measure the number density distribution of the rest equivalent width, $W_{0,\lambda2796}$ (${\rm dN}/{\rm d}W_{0,\lambda2796}$), around LRGs as shown in Figure~\ref{rew_dist}. Here we only use systems with quasar spectra $S/N>6$ to avoid incompleteness. 
To focus on absorbers predominantly linked with halos, we display only those within $r_p \leq 400$~kpc. 
Following \citet{Zhu2013b}, we adopt an exponential function, ${\rm dN}/{\rm d}W_{0,\lambda2796}~=~N^{*}~\times~e^{{-W_{0,\lambda2796}/W^{*}}}$, to fit the $W_{0,\lambda2796}$ number density distribution, and our best-fit parameters are shown in Table~\ref{rew_para}.
No significant difference is found between the radio LRGs and the control samples for both LoTSS and VLASS, with the KS test p-values 0.901 and 0.204\footnote{Here we used a nearly completed sample of sources with S/N $\geq 8$ for the KS test.}, respectively. 
We note that the shape of ${\rm dN}/{\rm d}W_{0,\lambda2796}$ around LRGs with $W^{*}\sim0.55\pm0.05$ is consistent with the shape of overall MgII population from \citet{Zhu2013b} with $W^{*}\simeq0.6$ at redshift $\sim 0.65$. 

\subsection{Gas kinematics} \label{losvelocity}
\begin{figure*}
\begin{center}
\includegraphics[width=0.8\linewidth]{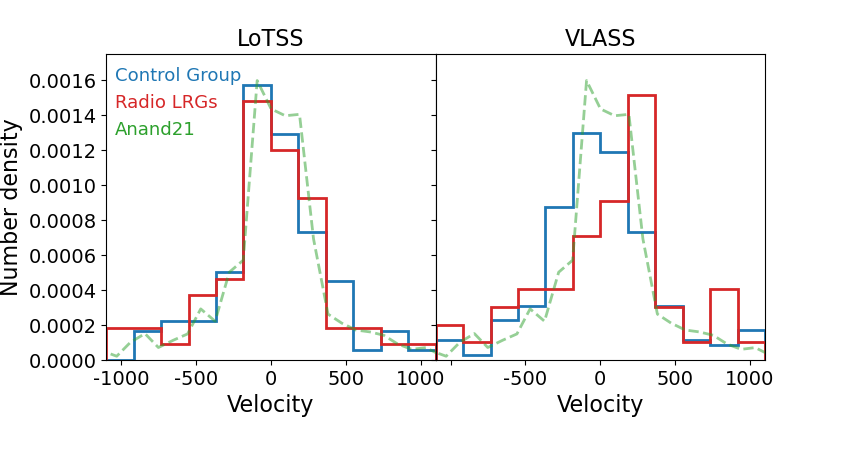}
\end{center}
\caption{Distribution of the light-of-sight gas-galaxy velocity difference ($dv$) of the LRG within $r_p\leq400$~kpc. The left panel displays the $dv$ distribution of LRGs from the LoTSS samples, and the right panel shows that of LRGs from the VLASS samples. The red and blue lines correspondingly represent the radio groups and control groups. We also show the distribution of passive galaxies in \citet{Anand2021}.} 
\label{velocity}
\end{figure*}

We now explore the gas kinematics of radio galaxies by investigating the line of sight (LoS) velocity difference, $dv$, between the gas traced by MgII absorption lines and the galaxies.  
The $dv$ distributions are shown in Figure~\ref{velocity}, where we consider absorbers mostly bounded with the halos within $r_p \leq 400$~kpc. 
To quantify the possible difference between the $dv$ distributions, we perform the KS test between the $dv$ distribution of radio LRGs and their control sample. The calculation yields p-values of 0.507 for the LoTSS sample and 0.061 for the VLASS sample respectively. This indicates that there is no detectable difference in $dv$ distributions between the radio galaxies and control groups from both LoTSS and VLASS. In addition, in Figure~\ref{velocity}, we show the $dv$ distribution from SDSS LRGs \citep{Anand2021}. The $dv$ distributions of the LoTSS samples are consitent with the SDSS LRGs while distributions of the VLASS samples are broader than the distribution of SDSS LRGs. We discuss the difference in Section~\ref{discussion}.

\begin{figure*}
\begin{center}
\includegraphics[width=0.8\linewidth]{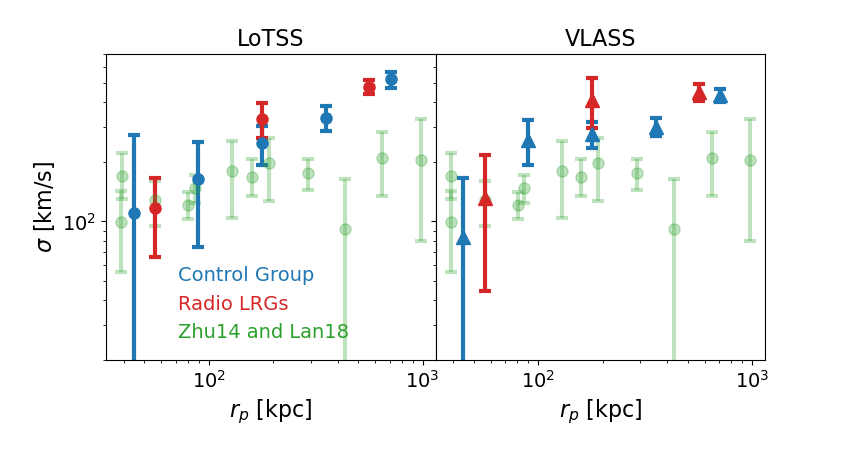}
\end{center}
\caption{Gas velocity dispersion as a function of impact parameter. The left and right panels show the results of LoTSS and VLASS LRGs respectively. The red, blue, and green points represent the radio LRGs, control groups, and the passive galaxies from \citet{Zhu2014, Lan2018}, respectively. The velocity dispersion obtained in this work is estimated using MAD scaled to the standard deviation.}
\label{vel_dis}
\end{figure*}

We further measure the LoS velocity dispersion as a function of $r_p$. 
For each $r_p$ bin, we use the median absolute deviation (MAD) scaled to the standard deviation as an estimator for the LoS velocity dispersion and estimate the corresponding errors by bootstrapping. 
We have subtracted the instrumental resolution ($\sim 40~{\rm km/s}$) and redshift uncertainty of LRGs ($\sim 50~{\rm km/s}$) in quadrature to measure the intrinsic velocity dispersion of the gas.
The results are illustrated in Figure~\ref{vel_dis}, and Table~\ref{vel_dis_rpbin} further shows the dispersion values for sources within $r_p \leq 400$~kpc and $r_p > 400$~kpc. 
We find that the measurements around radio LRGs and the control samples are consistent with each other. For the gas in the inner region ($r_p \leq 400~{\rm kpc}$), the radio LRGs and the control samples from both LoTSS and VLASS have a lower velocity dispersion ($\sigma \sim 200-300$ km/s), similarly to the results in previous studies \citep{Zhu2014, Lan2018}, while at larger scales, gas exhibits a higher velocity dispersion, reaching $\sim 500{\rm km/s}$. This behavior is consistent with the contribution from Hubble flow at larger scales.

\begin{table} [h!]
\setlength\tabcolsep{3pt}
\begin{center}
\caption{Gas velocity dispersion of our samples separated with two impact parameter bins.}
    \begin{tabular}{lcccc}
    \hline
    \hline
    & \multicolumn{2}{c}{$r_p \leq 400$~kpc} & \multicolumn{2}{c}{$r_p > 400$~kpc} \\      
      & LoTSS & VLASS  & LoTSS & VLASS \\ 
      & [{\rm km/s}]  & [{\rm km/s}] & [{\rm km/s}] & [{\rm km/s}] \\  
    \hline
    Radio LRGs&  $289\pm58$ & $313\pm72$ & $480\pm38$ & $445\pm45$  \\
    Control Group& $231\pm42$ & $279\pm27$ & $494\pm47$ & $426 \pm30$ \\
    \hline
    \end{tabular}
\label{vel_dis_rpbin}
\end{center}
\end{table}

\begin{figure}
\begin{center}
\includegraphics[width=0.9\linewidth]{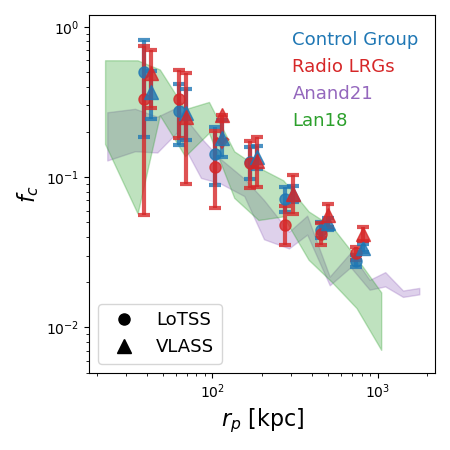}
\end{center}
\caption{Comparison of the covering fraction ($W_{0,\lambda2796}>0.4 \rm \, \AA$) with previous studies. The blue points, orange points, purple band, and green band are the $f_c$ for the LRGs from our LoTSS samples, our VLASS samples, \citet{Anand2021}, and \citet{Lan2018}, respectively. The data from \citet{Anand2021} and \citet{Lan2018} are scaled to an equivalent redshift of 0.7 using evolution trend from \citet{Lan2020}.}
\label{cov_compare}
\end{figure}

\section{Discussion} \label{discussion}

\subsection{Comparison with previous studies}
We compare our measurement of $f_c$ ($W_{0,\lambda2796}>0.4 \, \rm \AA$) of radio-detected LRGs with the measurements of SDSS LRGs from \citet{Lan2018} and \citet{Anand2021} in Figure~\ref{cov_compare}. 
Overall, the decreasing trends of $f_c$ are consistent with each other with similar slopes. As DESI and SDSS adopt different criteria for selecting LRG samples, which yield different stellar mass and redshift distributions, we scale the SDSS measurement using the evolution trend based on \citet{Lan2020}. 
If we consider the median redshifts of SDSS LRGs ($z\sim0.5$) and radio-detected LRGs ($z\sim0.7$) in this work, the redshift evolution of $f_c$ of strong absorbers will yield $\bigg(\frac{1+0.7}{1+0.5}\bigg)^{2.5}\sim1.4$
As illustrated in Figure~\ref{cov_compare}, after adjusting for evolution, the $f_c$ of the SDSS LRGs from \citet{Lan2018} and \citet{Anand2021} is consistent with that of the radio-detected LRGs from DESI at a given impact parameter. 

In addition to the covering fraction, we observe a larger gas velocity dispersion at larger scales, as compared to the measurement from \citet{Zhu2014} (Figure~\ref{vel_dis}). 
This difference is possibly due to the approaches used to calculate the velocity dispersion. \citet{Zhu2014} measure the velocity dispersion from the composite spectra via stacking analysis, Together with a median filtering for spectral processing, this method might mediate the signal far from the line center. We note that using a similar approach as adopted in this work, \citet{Anand2021} also find a larger velocity dispersion for gas of LRGs in the outer region. 

\begin{figure}
\begin{center}
\includegraphics[width=0.9\linewidth]{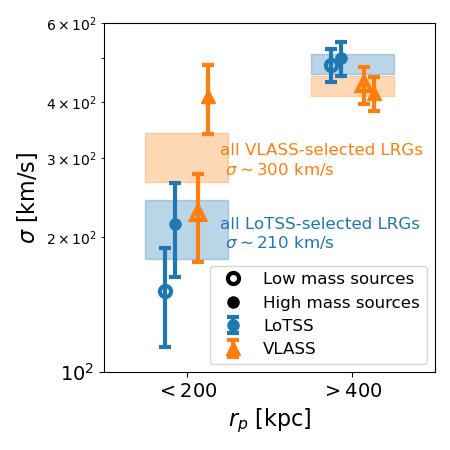}
\end{center}
\caption{Velocity dispersion as a function of stellar masses. The blue data represents both the radio LRGs and the corresponding control galaxies from the LoTSS catalog, while the orange data represents that from the VLASS catalog. The blue and orange bands show the dispersion for all sources selected with LoTSS and VLASS. The filled points are the results from high-mass sources, while the open points are those from low-mass sources. The term `low-mass sources' and `high-mass sources' refer to those with stellar mass above and below the median of the samples, which is $10^{11.2}~M_\odot$ for LoTSS and $10^{11.4}~M_\odot$ for VLASS.}
\label{vel_compare}
\end{figure}

Finally, among our own measurements, we quantify the potential differences between all the radio and the control LRGs selected with the LoTSS and the VLASS, regardless of whether it has radio emission or not. The gas velocity dispersion\footnote{Similar to Figure 9, the dispersion shown in this section is estimated using MAD scaled to the standard deviation.} of LRGs in the inner CGM ($<200$ kpc) for LoTSS samples is $211\pm 32~{\rm \, km/s}$, while that for the VLASS samples is $303\pm38~{\rm \, km/s}$, as illustrated by the blue and red band in Figure~\ref{vel_compare}. 
The inner gas around radio-detected LRGs and control LRGs from VLASS exhibits higher velocity dispersion compared to those from LoTSS. This difference in gas properties can be explained by the stellar mass difference of the two samples. 
In Figure~\ref{vel_compare}, we divide the sample based on their median stellar mass, $\sim 10^{11.2}~M_\odot$ and $\sim 10^{11.4}~M_\odot$ for LoTSS and VLASS, into high-mass and low-mass sources. The figure indicates that the gas dispersion for the absorbers around high-mass LRGs (stellar mass above the median value) from VLASS-selected samples is apparently higher than the others. Additionally, LRGs with comparable stellar masses ($\sim 10^{11}-10^{11.4}$), for LoTSS-selected sources and low-mass VLASS-selected sources, display similar gas dispersion in their CGM.
We note that the absorbers beyond $r_p=400$ are affected by the haloes of other nearby galaxies, and their velocity dispersion is not strongly correlated with the stellar mass. 

As shown in Figure~\ref{stellar}, the median stellar mass of the VLASS samples is $\sim 0.2$~dex higher than that of the LoTSS samples. 
This 0.2~dex difference in the median stellar mass yields a larger difference in halo mass at the high mass end.
Based on the stellar-to-halo mass relation (SHMR) from \citet{Girelli2020}, the estimated halo mass difference between the two samples can reach $\sim 0.5$~dex. If we assume the typical halo mass of the LoTSS sample is similar to the overall DESI LRG population $10^{13.4} \, M_{\odot}$ \citep[e.g.,][]{YuanLRGs}, the typical halo mass of the VLASS sample will be about $10^{13.9} \, M_{\odot}$. Their corresponding dark matter velocity dispersion values are $\sim 270$ and $\sim 400$ km/s based on \citet{ElahiVD}. This yields a similar sub-virial gas motion $\frac{\sigma_{gas}}{\sigma_{DM}}\sim 0.75$ within 200 kpc around VLASS and LoTSS samples. We note that this value is higher than previous measurements around SDSS LRGs ($\frac{\sigma_{gas}}{\sigma_{DM}}\sim 0.5$). We will explore such a difference in the future with the entire DESI LRG sample. The differences in gas kinematics illustrated in Figure~\ref{velocity} between LoTSS-selected and VLASS-selected sources could also be influenced by the disparities in stellar masses of the two samples. Additionally, the properties of the VLASS samples are consistent with the measurements of galaxy clusters, showing that the CGM in clusters traced by MgII has a higher covering fraction and velocity dispersion \citep[e.g.,][]{Cherrey2023, Anand2022}.

\subsection{CGM mass around radio galaxies} 
With the covering fraction obtained in Section~\ref{covering}, we can quantify the typical amount of cool HI gas around the radio-detected LRGs. We estimate the mass of neutral hydrogen traced by MgII absorbers residing in the CGM with 
\begin{equation}
    M_{\rm HI}(<400 \, \rm kpc) \sim 2\pi\, m_{\rm H} \int_{10 \rm \, kpc}^{400\rm \, kpc} 	\hat{N}_{\rm HI}\, f_{c}(r_{p})r_{p}dr_{p},
\end{equation}
where $\hat{N}_{\rm HI}$ is the empirical relation between $W_{0,\lambda2796}$ and $N_{HI}$ from \citet{Lan17},
\begin{equation}
    \hat{N}_{\rm HI} = 10^{18.96}\left(\frac{W_{0,\lambda2796}}{1 \,  \rm \AA}\right)^{1.69}(1+z)^{1.88} \, \rm cm^{-2}.
\end{equation}
The median value of samples $z\sim0.67$ is adopted. 
For weak absorbers, we take the median value of our MgII absorbers $W_{0,\lambda2796}\sim 0.65 \rm \,\AA$ and for strong absorbers, we take the median value $W_{0,\lambda2796}\sim 1.5 \,\rm \AA$. The final estimated $M_{HI}$ for LRGs with LoTSS is $9.9\times10^{9}~{\rm M_\odot}$ and for LRGs with VLASS is $1.3\times10^{10}~{\rm M_\odot}$. Given that the covering fractions around control samples are consistent with the radio LRGs, the control samples have similar mass of neutral hydrogen within 400 kpc. The excess of gas may originate from thermal instabilities in the halos of LRGs \citep{Huang2016, Nelson2020}, star-forming satellite galaxies within the LRG halos and their neighboring halos \citep{Hafen2019, Lan2020}, as well as gas accreting through filaments \citep{Kerevs2009, Huang2016}. 

This result indicates that there is a non-negligible amount of cool neutral gas around massive galaxies with possible radio-mode feedback in action. 
This mass measurement can place constraints on the models of feedback in simulations. For example, \citet{Khrykinradio_simulation} use the SIMBA simulations \citep{Simba} to explore the impact of various feedback models on the CGM properties and find that including radio-mode feedback reduces the amount of baryons in the CGM by a factor of 5 in halo mass $\sim 10^{13} \, M_{\odot}$, while only reducing by a factor of 2 as the halo mass increasing to $\sim 10^{13.5} \, M_{\odot}$. For another example, \citet{TNGradiofeedback} show that radio-mode feedback changes the thermal properties of the CGM by enhancing the entropy and thereby increasing the gas cooling time. The properties of the cool gas are expected to be affected as well. 
As summarized in \citet{ARAA_simulations}, adopting different assumptions and subgrid recipes of AGN feedback, the state-of-the-art simulations predict different amount of baryons in the halos. 
From these simulations, one can estimate the amount of cool neutral gas around radio galaxies, compare with the measurements reported in this work, and test those feedback models. 
Furthermore, our gas dispersion measurements could also constrain the gas outflow and their initial velocity for the feedback models \citep{Nelson2019, Lan2019, Mitchell2020}.

\subsection{Impact of the radio jets} \label{radiojet}
From the X-ray and radio observations showing the interaction between radio jets and hot gas, one would expect that as they propagate, radio jets impact the surrounding medium.  
However, our results show no discernible differences between the properties of the cool CGM of galaxies with and without radio emission. Unlike the hot CGM, radio-mode feedback might not produce detectable effects on the properties of the cool CGM with our current sample. In the following, we discuss two possible scenarios for such a result. 

\textbf{Time and Distance Scales:} The majority of our radio sources do not have an extended radio emission structure. It is possible that the radio-mode feedback just starts its operation. Therefore, the radio jets do not have sufficient time to propagate into larger scale ($\sim 100\, \rm kpc$) \citep[e.g,][]{Hardcastle2019}. 
To comprehensively analyze gas properties for galaxies with various radio morphologies, more extensive radio surveys are required. These surveys should feature resolution and sensitivity sufficient to distinguish small extended structures and detect low surface brightness sources. 

Another possibility is that the power of feedback and/or the duration of feedback is not strong and long enough to propagate into larger scales. For radio galaxies with stellar masses around or above $10^{11}~{\rm M_\odot}$, their typical active period spans 10-100 Myr \citep[e.g.,][]{McNamara2012, Turner2015} and their jets can on average span tens of kpc, with the rare bright ones possibly extending to $\sim 200-300$ kpc or even larger \citep[e.g.,][]{Turner2015, Hardcastle2019, Lan2021}. Given that we only have a handful of sightlines probing the inner CGM, the effect of radio-mode feedback might be below the noise level of the current measurements.

\textbf{Opening angle:} While powerful radio jets can extend to several hundred kpc, the typical observed jet structures are asymmetric with certain direction \citep[e.g.,][]{Hardcastle2020}, being consistent with the unified model of AGNs \citep{Urry1995, Netzer2015, Padovani2017}. For example, based on the observations from the Monitoring Of Jets in Active galactic nuclei with VLBA Experiments (MOJAVE) program\footnote{\url{https://www.cv.nrao.edu/MOJAVE/index.html}}, the radio jets typically have relatively small opening angles with a median value of around $20^\circ$ \citep{Pushkarev2017}.
Given the small opening angles, the background lines of sight may not intercept with the most affected regions by the radio jets. This will dilute the signals and prevent us from detecting the impact of radio jets with a limited number of sightlines in the inner CGM. 

DESI is going to observe $\sim 8$ million LRGs \citep{DESI_LRG} and $\sim3$ million QSOs \citep{DESI_QSO} during its five-year mission. In its first year of the main survey, DESI has already observed $\sim 3.5$ million LRGs and $\sim 1.5$ million QSOs. It is anticipated that the number of known LRG-QSO pairs will triple after DESI completes its operation, with the number of their radio counterparts also expected to increase by approximately threefold. With such a large number of LRG-QSO pairs, we will be able to conduct a more comprehensive study of the relationship between radio jets and the CGM of LRGs.

\section{Conclusions} \label{conclusion}
We investigate how radio-mode feedback impacts the properties of the cool CGM traced by MgII absorbers.
To this end, we constructed two large samples of $\sim 30,000$ radio LRGs and background QSO pairs using the latest DESI spectroscopic measurements and two large radio source catalogs from LoTSS and VLASS. We also built two corresponding control galaxy samples that match the host galaxy properties of the radio LRGs but without detected radio emission. With these datasets, we measured and compared the CGM properties around radio LRGs and their control samples and explored possible signals correlating with the presence of radio emission. 
 
Our results show no significant differences between the properties of the cool CGM of radio LRGs and their control samples, including the gas radial distribution, MgII rest equivalent width distributions and the gas kinematics. This result indicates that there is no detectable correlation between the presence of radio emission and the cool CGM properties. 
Among the LRG samples, we find that the gas velocity dispersion around radio LRGs in VLASS is higher than that around radio LRGs in LoTSS. This can be explained by the fact that radio LRGs in VLASS on average have higher stellar masses and therefore prefer to live in more massive halos. 
Finally, we estimate the amount of cool gas mass in the CGM of radio LRGs being approximately $10^{10} M_{\odot}$ --- a none-negligible amount of cool gas reservoir. 
These novel measurements can be used as strong observational constraints on the models of AGN feedback adopted in simulations, providing valuable insights into the impact of radio-mode feedback on baryon distribution around galaxies and the growth of massive galaxies. 

In the near future, combining the updated versions of the LoTSS and VLASS catalogs, as well as upcoming deeper and wider sky surveys from the Square Kilometre Array \citep[SKA,][]{SKA} and the next generation VLA \citep[ngVLA,][]{ngVLA} with optical large spectroscopic datasets, including the complete 5-year DESI dataset, we will be able to increase the sample size by at least a factor of 2 and obtain more precise measurements of the CGM properties of radio galaxies. 
With such measurements, one will possibly reveal the signatures of radio-model feedback on the cool CGM for the first time and thereby test the current models of AGN feedback and galaxy evolution.



\section*{Data Availability}
All data points shown in the figures are available in a machine-readable form on Zenodo \url{https://doi.org/10.5281/zenodo.11143902}.

\section*{Ackowledgments}
We thank Vicky Fawcett and Siwei Zou for their constructive comments that greatly helped improve the paper. YLC and TWL were supported by the National Science and Technology Council (MOST 111-2112-M-002-015-MY3), the Ministry of Education, Taiwan (MOE Yushan Young Scholar grant NTU-110VV007, NTU-110VV007-2, NTU-110VV007-3), National Taiwan University research grant (NTU-CC-111L894806, NTU-CC-112L894806, NTU-CC-113L894806). YLC is supported by NSTC 112-2811-M-002-059.

This material is based upon work supported by the U.S. Department of Energy (DOE), Office of Science, Office of High-Energy Physics, under Contract No. DE–AC02–05CH11231, and by the National Energy Research Scientific Computing Center, a DOE Office of Science User Facility under the same contract. Additional support for DESI was provided by the U.S. National Science Foundation (NSF), Division of Astronomical Sciences under Contract No. AST-0950945 to the NSF’s National Optical-Infrared Astronomy Research Laboratory; the Science and Technology Facilities Council of the United Kingdom; the Gordon and Betty Moore Foundation; the Heising-Simons Foundation; the French Alternative Energies and Atomic Energy Commission (CEA); the National Council of Science and Technology of Mexico (CONACYT); the Ministry of Science and Innovation of Spain (MICINN), and by the DESI Member Institutions: \url{https://www.desi.lbl.gov/collaborating-institutions}. Any opinions, findings, and conclusions or recommendations expressed in this material are those of the author(s) and do not necessarily reflect the views of the U. S. National Science Foundation, the U. S. Department of Energy, or any of the listed funding agencies.

The authors are honored to be permitted to conduct scientific research on Iolkam Du’ag (Kitt Peak), a mountain with particular significance to the Tohono O’odham Nation.

LOFAR data products were provided by the LOFAR Surveys Key Science project (LSKSP; https://lofar-surveys.org/) and were derived from observations with the International LOFAR Telescope (ILT). LOFAR (van Haarlem et al. 2013) is the Low Frequency Array designed and constructed by ASTRON. It has observing, data processing, and data storage facilities in several countries, which are owned by various parties (each with their own funding sources), and which are collectively operated by the ILT foundation under a joint scientific policy. The efforts of the LSKSP have benefited from funding from the European Research Council, NOVA, NWO, CNRS-INSU, the SURF Co-operative, the UK Science and Technology Funding Council and the Jülich Supercomputing Centre.

The National Radio Astronomy Observatory is a facility of the National Science Foundation operated under cooperative agreement by Associated Universities, Inc. CIRADA is funded by a grant from the Canada Foundation for Innovation 2017 Innovation Fund (Project 35999), as well as by the Provinces of Ontario, British Columbia, Alberta, Manitoba and Quebec.







\appendix

\section{Detection Rate Simulation} \label{detection}
\begin{figure}
\begin{center}
\includegraphics[width=0.9\linewidth]{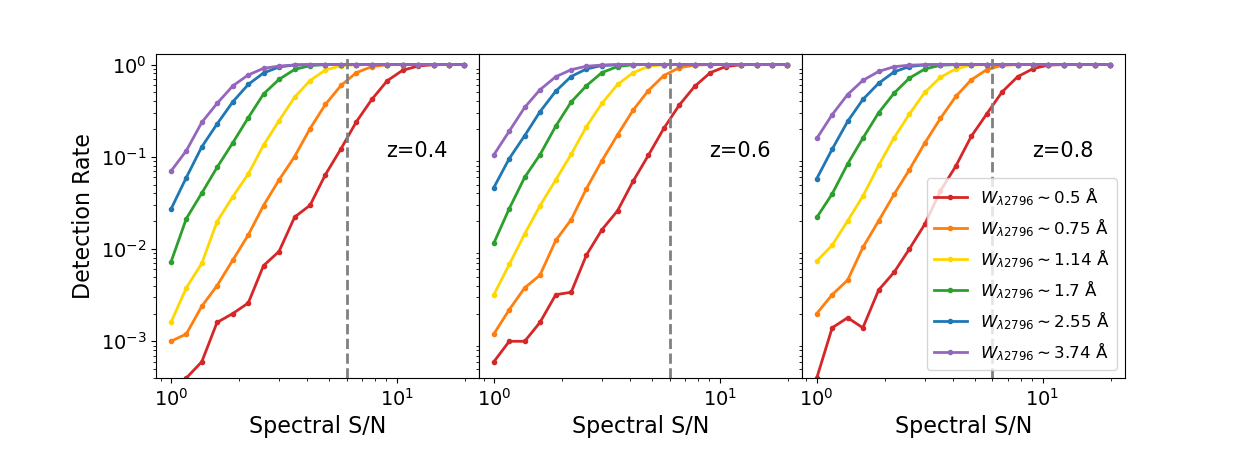}
\end{center}
\caption{Detection rate of MgII absorbers as a function of absorber redshift, $W_{0,\lambda2796}$, and S/N of the spectra. The left to right panels show the results for redshift bins 0.4, 0.6, and 0.8, respectively. The different colors represent various rest equivalent widths for the mock spectra. The vertical dashed lines represent the spectral S/N cut for weak absorbers ($0.4<W_{0,\lambda2796}<1 \rm \, \AA$). }
\label{detect_rate}
\end{figure}

We simulated the detection rate using mock absorption lines to calibrate the number of detected weak absorbers found in realistic QSO spectra with low S/N. Figure~\ref{detect_rate} illustrates the detection rate of MgII absorbers as a function of absorber redshift, $W_{0,\lambda2796}$, and S/N of the spectra. More details about the completeness of our absorption samples are shown in section~\ref{complete}. 

\bibliography{sample631}{}
\bibliographystyle{aasjournal}



\end{document}